# Integrated electro-optic digital-to-analog link for efficient computing and arbitrary waveform generation

Yunxiang Song[1,2,*,†], Yaowen Hu[1,3,*,†], Xinrui Zhu[1], Keith Powell[1], Letícia Magalhães[1], Fan Ye[4], Hana Warner[1], Shengyuan Lu[1], Xudong Li[1], Dylan Renaud[1,4], Norman Lippok[1,5], Di Zhu[6,7,8], Benjamin Vakoc[5], Mian Zhang[4], Neil Sinclair[1], Marko Lončar[1,†]

[1] John A. Paulson School of Engineering and Applied Sciences, Harvard University, Cambridge, MA 02138, USA
[2] Quantum Science and Engineering, Harvard University, Cambridge, MA 02138, USA
[3] State Key Laboratory for Mesoscopic Physics and Frontiers Science Center for Nano-optoelectronics, School of Physics, Peking University, Beijing 100871, China
[4] HyperLight Corporation, 501 Massachusetts Ave, Cambridge, MA 02139, USA
[5] Wellman Center of Photomedicine, Harvard Medical School and Massachusetts General Hospital, Boston 02114, USA
[6] Department of Materials Science and Engineering, National University of Singapore, Singapore 117575, Singapore
[7] Centre for Quantum Technologies, National University of Singapore, Singapore, 117543, Singapore
[8] A*STAR Quantum Innovation Centre (Q.InC), Institute of Materials Research and Engineering (IMRE), Agency for Science, Technology and Research (A*STAR), Singapore 138634, Singapore
[*] These authors contributed equally
[†] ysong1@g.harvard.edu, yaowenhu@pku.edu.cn, loncar@seas.harvard.edu

**The rapid growth in artificial intelligence and modern communication systems demands innovative solutions for increased computational power and advanced signaling capabilities. Integrated photonics, leveraging the analog nature of electromagnetic waves at the chip scale, offers a promising complement to approaches based on digital electronics. To fully unlock their potential as analog processors, establishing a common technological base between conventional digital electronic systems and analog photonics is imperative to building next-generation computing and communications hardware. However, the absence of an efficient interface has critically challenged comprehensive demonstrations of analog advantage thus far, with the scalability, speed, and energy consumption as primary bottlenecks. Here, we address this challenge and demonstrate a general electro-optic digital-to-analog link (EO-DiAL) enabled by foundry-based lithium niobate nanophotonics. Using purely digital inputs, we achieve on-demand generation of (i) optical and (ii) electronic waveforms at information rates up to 186 Gbit/s. The former addresses the digital-to-analog electro-optic conversion challenge in photonic computing, showcasing high-fidelity MNIST encoding while consuming 0.058 pJ/bit. The latter enables a pulse-shaping-free microwave arbitrary waveform generation method with ultrabroadband tunable delay and gain. Our results pave the way for efficient and compact digital-to-analog conversion paradigms enabled by integrated photonics and underscore the transformative impact analog photonic hardware may have on various applications, such as computing, optical interconnects, and high-speed ranging.**

## Introduction

The development of efficient, high-bandwidth communication systems and novel computational paradigms has been driven by advancements in high-performance computing and complex signal processing infrastructure, leading to substantial scientific and industrial impact[1,2]. Examples include tensor processing units[3], neuromorphic chips[4], and high-frequency electronics[5,6], critically satisfying the most stringent demands in compute power and data transmission rates. Recent years have also seen a burgeoning interest in developing analog hardware to handle exponentially rising data volumes and generate ultrabroadband electronic waveforms[7]. Photonics is ideally suited for these tasks, offering higher capacity and throughput than incumbent electronic approaches, while also consuming substantially less energy[8–16]. Moreover, photonic integration supports compact form factors, offers excellent scalability, and enables cost-effective production.

For photonic technologies to be seamlessly integrated into computing and communication systems, a high-speed, energy-efficient interface between the digital electronic domain and the analog photonics domain is essential, where the former is used for information storage and the latter for photonics-accelerated information processing and transfer. Currently, this interface is primarily established using a two-step process: digital-to-analog converters (DACs) are used to convert binary words into analog electrical signals, which are then used to drive electro-optic (EO) modulators. However, there is a significant system complexity and cost associated with high-speed DAC circuits that are needed to drive state-of-the-art EO modulators to their analog bandwidth limits exceeding one-hundred gigahertz (GHz)[17,18]. To address this bottleneck, prior efforts have utilized silicon photonics to integrate DAC and EO modulation for higher-order modulation format optical communication[19–25]. However, the silicon platform and its plasma-dispersion-based electro-optic conversion are associated with large insertion losses, bandwidth limitations, and high reverse-bias/AC-driving, limiting these systems to only two-bit resolution and prohibited efficient performance. A general, scalable digital-to-analog link that meets the critical requirements for high-performance, large-scale analog photonic computing and signaling remains elusive.

Here, we overcome these limitations by utilizing thin-film lithium niobate (TFLN) photonics to demonstrate an electro-optic digital-to-analog link (EO-DiAL) device, based on a multi-port Mach-Zehnder interferometer (MZI). The device can realize high-speed, multi-level optical encoding using two-level electrical drives, fully leveraging both efficient electro-optical interaction[26–28] and low-optical and microwave loss properties[29,30] of the TFLN platform. The EO-DiAL acts as an alternative to conventional electronic DAC and EO modulator pairs, with potential to become a key component in next-generation photonic computing architectures as they heavily rely on EO conversion[31]. Further, it may advance and broaden the microwave-photonic signal generation toolset by offering a novel radiofrequency arbitrary waveform generation (RF-AWG) method. Considering these two application spaces, we demonstrate that the EO-DiAL produces a data stream in the optical domain with 4 bit amplitude- and 21.505 ps-temporal resolution when it is driven by 4-bit binary words, leading to an effective data rate of 186 Gb/s and energy consumption of 0.058 pJ/bit. To illustrate the potential of this platform, we encode onto the optical carrier down-sampled Modified National Institute of Standards and Technology (MNIST) handwritten digits with high fidelity. We then combine the EO-DiAL optical output with fast photodetection to perform optical to electronic (OE) conversion, thus realizing an electronic RF-AWG. In contrast to conventional microwave photonic methods for AWG such as optical pulse shaping[32–35], our approach is based on continuous-wave light inputs, does not require optical fanout or resonant filters, and is facilitated by coherent phase control in the EO-DiAL. Hence, the synthesized arbitrary waveforms exhibit no signal downtime, and they are compatible with RF-over-fiber (RFoF) link technologies[36], among which we show tunable RF delay and measurements towards broadband power gain in the generated waveforms.

## Results

### EO-DiAL concept

The EO-DiAL concept generalized to $N$ bits of resolution, utilizing a segmented MZI structure with traveling wave electrodes on a chip, is illustrated in Fig. 1a. An $N$-bit EO-DiAL features $N$ pairs of push-pull electrode segments, instead of the usual one pair, each twice longer than the next one and thus requiring two-times smaller voltage ($V_\pi$) to change the phase of the propagating light by $\pi$ (180 degrees). In other words, if the longest segment is characterized by switching voltage $V_\pi$, then the second longest segment by $2V_\pi$, and the shortest segment by $2^{N-1}V_\pi$. A stream of $N$-bit binary words to be optically encoded are represented by $N$ voltage signals. Words are encoded sequentially in time, each represented by the set of $V_n(t = m \cdot T)$. Here, $n$ is the segment number ($n = 0$ being the shortest segment with electrode length $L$, corresponding to the least significant bit and $n = N - 1$ the longest segment with electrode length $2^{N-1}L$, corresponding to the most significant bit), $1/T$ is the bit rate, $m$ is an integer index for the words, and voltage $V_n$ assumes one of the two values ($V_{high}, V_{low}$). When all segments are driven together by a set of $V_n(t)$, the optical fields in the push and pull arms accumulate a phase difference $\Delta\phi$ given by:

$$\Delta\phi(t) = \frac{\pi}{2^{N-1}V_\pi}\left(\sum_{n=0}^{N-1} V_n(t)2^n\right). \tag{1}$$

The power transfer function of the EO-DiAL is given by $T = \frac{1}{2}(1 + \sin(\Delta\phi(t)))$, which is proportional to $\Delta\phi(t)$ in the limit of small $\Delta\phi(t)$ (satisfied when $V_n(t) \ll V_\pi$) and when the MZI is biased at quadrature. Since purely digital modulations partition the total $\Delta\phi(t)$ into $2^N$ levels (quantity in parenthesis of Eq. 1), the EO-DiAL output power is $N$-bit-resolved, and both EO conversion and digital-to analog conversion is achieved simultaneously, without an external electronic DAC. The 4-bit device implementation and corresponding bit-wise addressable transmission is explicitly presented in Fig. 1b.

**TFLN implementation**

The EO-DiAL device, shown in Fig. 2a, occupies a compact footprint of only 30.73 mm² and is fabricated using a foundry-based wafer-scale TFLN process (Multi-Project Wafer run, HyperLight Corporation). Main components of the device are shown in Fig. 2b and include electrode pads, capacitance-loaded coplanar lines, 50 Ω terminations, metal routing, low-loss fiber-to-chip grating couplers, and heaters (see Supplementary information section for details about these and other components). In particular, four push-pull electrode segments (Fig. 2c S0-S3, with 0.25, 0.50. 1.00, and 2.00 cm lengths, respectively) form an MZI, mapping to four bits of resolution. Note that low optical loss and wide EO bandwidth, characteristic of TFLN modulators, ensure fast and low energy consumption of the EO-DiAL (see Supplementary information for optical loss characterizations). To operate the EO-DiAL with large bandwidth, the frequency-dependent $V_\pi$s for the electrode segments must satisfy the factor-of-two relationship throughout the complete frequency content of the digital modulation patterns. This translates to a flat frequency response requirement for each segment, justifying the need for capacitance-loaded RF transmission lines which minimize $V_\pi$-bandwidth tradeoffs[37]. The measured $V_\pi$ for segments S0-S3 are given by 10.19, 5.97, 3.27, and 1.69 V, respectively (Fig. 2d). To account for residual deviations from the designed $V_\pi$ relationship, likely due to fabrication imperfection, the amplitudes of $V_n(t)$ defined as $|V_n|$ are slightly tuned to enforce linearity between $|V_n|/V_\pi$ and segment length $L_n$ (Fig. 2e), resulting in $|V_n|$ of 200, 195, 210, and 215 mV, respectively. This set of $|V_n|$ is characteristic of the device and is fixed throughout the experiments. Without these imperfections, $|V_n|$ should be identical for all $n$. Normalized microwave S-parameters (Fig. 2f) for segments S0-S2 indicate low electrical-electrical (EE) return loss and near flat EO response up to 50 GHz modulation frequencies (limited by the bandwidth of our measurement equipment). On the other hand, segment S3 has an EO 3 dB cutoff of about 30 GHz and a simulated 5 dB cutoff of about 100 GHz (limited by microwave loss), which, though sufficient for this work, may be improved in future iterations. Finally, the EO-DiAL linearity is examined by synchronizing four binary modulation sequences (each sampled at 46.5 Gb/s) such that, when combined, probe each level of the device's four bits of amplitude resolution. The amplitudes conform well to a linear fit, suggesting high degree of linearity (Fig. 2g).

**Efficient EO digital-to-analog interface**

To benchmark the EO-DiAL's performance for efficient EO conversion within photonic computing architectures (Fig. 3a), two key performance metrics are systematically evaluated, namely the maximal encoding speed and the minimal energy requirement for high fidelity encoding. First, a set of four binary modulation sequences clocked at an aggregate data rate of 124 Gb/s and 186 Gb/s are delivered to modulation segments S0-S3 of the EO-DiAL. Each sequence consists of 500 random bits that altogether combine into one sequence of 500 random 4-bit words. The 4-bit-resolved amplitude is directly observed on a real-time oscilloscope (RTO) to evaluate encoding quality. However, in real computing applications, this encoded light would be routed into a photonic processor. As shown in Fig. 3b, the optical amplitude closely follows the ideal random data sequence at both data rates. Thus, the electronic data, presented in binary form, is successfully encoded onto the optical carrier. We note that imperfections at 186 Gb/s manifest as reduced amplitude swing and perceivable rise and drop times between words, evidenced by the zoom-in plots of Fig. 3b. At this highest data rate, deviations from perfect encoding are attributed to operating near the bandwidth limit of both the driving electronic circuit (digital driver and RF cable) and the photonic circuit (limited by segment S3). Indeed, the consequence of finite bandwidth on encoding practical data is shown in Fig. 3c. There, a down-sampled 28 by 28 MNIST image, represented by a sequence of 784 4-bit words, is converted into the optical domain at a rate of 186 Gb/s while consuming 0.104 pJ/bit. The primary features of the digit are accurately preserved. As the EO-DiAL operates as a DAC-free EO converter, we next explore the potential for machine learning algorithms to process these images. We apply a classification model on both computer-encoded and EO-DiAL-encoded test images. This provides insight into the potential performance of photonic computing systems implementing the model. We define the encoding accuracy of the EO-DiAL according to the overlap (represented by a confusion matrix) between both sets of classifications (Fig. 3d), and we find that 95% of 100 MNIST images are accurately encoded in the optical domain. Finally, the energy consumption of digital-to-analog and EO conversion is explored at the highest 186 Gb/s encoding rate, in relation to encoding fidelity. In Fig. 3e, reconstructed images at three different optical powers (-17.5, -11.6, and -5.6 dBm detected optical powers) and two different electrical powers ($|V_n|$=100, 200 mV applied driving voltages) suggest increases in the overall signal-to-noise ratio and dynamic range, when more power is consumed. A $|V_n|$~200 mV is sufficient to reach greater than 95% encoding accuracy across all optical power levels considered. On the other hand, $|V_n|$~100 mV with -17.5 dBm of detected optical power can still achieve 89% encoding accuracy, expending a total of 0.058 pJ/bit (Fig. 3f).

**Radiofrequency arbitrary waveform generation and manipulation**

A schematic of the EO-DiAL configured for RF-AWG is shown in Fig. 4a. Here, CW light is shaped by four digitally modulated segments S0-S3, such that an RF arbitrary waveform with 4-bit amplitude resolution is imprinted onto the optical carrier and subsequently undergoes OE conversion via fast photodetection. First, to demonstrate basic RF-AWG functionality, six standard waveforms (sine, trapezoid, sinc, chirp, square, and ramp) were separately synthesized (Fig. 4b). Next, a random arbitrary waveform with distinctly resolved features every $T$=21.505 ps were produced at eleven instances precisely 21.505 ps apart, by fine-tuning the optical delay with a motorized delay line. Matching modulation patterns in the detected waveforms showcase accurate word-by-word delay (Fig. 4c). Similarly, a sinc pulse was stored with high fidelity and low loss in optical fiber, released 480 ns later (Fig. 4d). Optical path length thus grants our EO-DiAL-based RF-AWG additional functionalities such as ps-precision, broadband RF delay and long-time RF signal storage and release. These functionalities are reminiscent of skew control between channels of an electronic RF-AWG and synchronization between multiple units within an electronic RF-AWG network, nominally requiring additional complex circuitry.

Next, broadband OE gain transfer of the RF-AWG is measured through the EO system response (Fig. 4e). A net positive gain up to 35 GHz can be achieved, after applying 11.6 dB of optical gain to amplify the EO-DiAL output supplied by an erbium-doped fiber preamplifier, in conjunction with transimpedance gain. The $S_{21}$ response of the system has identical shape to the transimpedance amplifier response, indicating flat optical to RF gain transfer. The 0 dB level is referenced to the summed VNA source power to drive segments S0-S3 (-6.0 dBm total, as described in the Methods section). Note that this response represents digital-to-analog response, in contrast to conventional RFoF link gain where the modulation and demodulated waveforms are identical. Finally, using a fast photodiode with high saturation power, the incident average optical power is varied from 1 to 15.9 mW, and the generated RF power of triangular waves (Fig. 4f inset) are estimated from the root-mean-squared voltage produced across a 50 Ω-load. Note that in this approach, broadband electronic gain on the synthesized waveform is inherited from the optical gain alone. Thus, broadband microwave gain can be straightforwardly achieved by leveraging the optical domain, in contrast to electronically amplifying electrical DACs which have electronic bandwidth limitations. As shown in Fig. 4f, an RF power up to 0.16 mW is generated, which is yet to surpass the total digital modulation power. Nevertheless, agreement between the measurement and an expected quadratic relation, between optical power incident on the photodiode and generated RF power, indicates linear OE conversion by the photodiode. This suggests that lower $V_\pi$ and higher photodiode saturation render above 0 dB digital-to-analog gain within reach (see Supplementary information for detailed discussion).

## Conclusion and Outlook

In summary, we presented a photonic-integrated EO-DiAL architecture that enables direct electrical-to-optical and digital-to-analog conversion. We utilized this device to demonstrate two practical use cases: an efficient EO converter suitable for photonic computing accelerators and a novel microwave-photonic arbitrary waveform generator. Notably, we utilized a foundry-based process for TFLN to demonstrate scalable and consistent fabrication of compact TFLN circuits. This process also highlights the promise for microwave/optical-circuit codesign at the foundry level for synergistic photonic-electronic structures on this platform. Currently, our device features an average $V_\pi \cdot L \sim 3.05$ V·cm, which is consistent with that expected from the foundry process design kit. State-of-the-art modulation efficiencies are 2.4 V · cm at telecommunication wavelengths while supporting EO 3 dB bandwidths exceeding 110 GHz[18,37], and much lower $V_\pi \cdot L$ are achievable at visible/near-infrared wavelengths featuring tens of GHz bandwidths[38,39]. Further improvements in modulator design and selection of optical carrier frequency, could substantially reduce electrical energy consumption and facilitate photonic-electric integration with high-speed CMOS pulse pattern generators[5]. Optical energy consumption can also be reduced by optimizing fiber-to-chip coupling and propagation loss[40] in the foundry process. Altogether, binary modulations with tens of millivolts swing and lasers consuming less than a milliwatt may be all that is needed for analog data encoding in photonic computing applications, while achieving encoding fidelities comparable to those in this work. In RF-AWG applications, better EO efficiencies directly induces greater slope efficiency and DAC gain (scaling inverse-quadratically with $V_\pi$)[36]. At the system-level, a higher degree of integration using chip-scale lasers[41–45] and on-chip fast photodiodes[46] can lead to more compact systems and achieve similar performance improvements as previously described, due to enhanced lasing efficiencies[41,42], reduced coupling losses[44], and greater detector responsivities[46]. Finally, further increasing the bit resolution is within reach by incorporating more modulation segments and by moderately scaling the drive voltage. Although, we note that achieving arbitrarily high bit resolution is not practically feasible due to the limited signal-to-noise ratio and the finite dynamic range of the EO conversion process. However, it is also not necessary[47]: the advantages of analog photonic hardware are often realized in applications where high bit precision is not critical (such as computing), and in many cases, limited precision can even enhance performance without compromising efficacy[8,12,31,48–50].

Given the speed and energy consumption assessments of the EO-DiAL, it may already substitute the essential EO modulator and electronic digital-to-analog converter in the coherent photonic computing framework[51,52] based on large-scale TFLN circuits[53–55]. It is worth noting that multiple continuous-wave (CW) tones, such as from a comb source[56,57], can simultaneously be utilized for parallelized data conversion across different wavelengths in a single pass of the device, common in photonic convolution acceleration[47,58]. In terms of offering a competitive yet more manageable pathway for RF-AWG, the EO-DiAL may be combined with on-chip optical amplification[59] and tunable delays facilitated by microring all-pass filter arrays[60] to enhance the compactness of our preliminary demonstrations. Altogether, our results suggest that the continued development of novel analog photonic hardware, such as the EO-DiAL, will broaden the design space for next-generation photonic computing, microwave-photonic RF-AWG, and other technologies such as high-capacity wireless/fiber-optic communications[61–64], as well as microwave/mmWave-photonic radar sensing[65,66] and signal processing[67,68].

**References**

1. LeCun, Y., Bengio, Y. & Hinton, G. Deep learning. *Nature* **521**, 436–444 (2015).

2. Winzer, P. J., Neilson, D. T. & Chraplyvy, A. R. Fiber-optic transmission and networking: the previous 20 and the next 20 years. *Opt. Express* **26**, 24190 (2018).

3. Graves, A. *et al.* Hybrid computing using a neural network with dynamic external memory. *Nature* **538**, 471–476 (2016).

4. Akopyan, F. *et al.* TrueNorth: Design and Tool Flow of a 65 mW 1 Million Neuron Programmable Neurosynaptic Chip. *IEEE Trans. Comput.-Aided Des. Integr. Circuits Syst.* **34**, 1537–1557 (2015).

5. Chen, X. *et al.* All-Electronic 100-GHz Bandwidth Digital-to-Analog Converter Generating PAM Signals up to 190 GBaud. *J. Light. Technol.* **35**, 411–417 (2017).

6. Song, H.-J. & Lee, N. Terahertz Communications: Challenges in the Next Decade. *IEEE Trans. Terahertz Sci. Technol.* **12**, 105–117 (2022).

7. Miller, D. A. B. Attojoule Optoelectronics for Low-Energy Information Processing and Communications. *J. Light. Technol.* **35**, 346–396 (2017).

8. Wetzstein, G. *et al.* Inference in artificial intelligence with deep optics and photonics. *Nature* **588**, 39–47 (2020).

9. Shastri, B. J. *et al.* Photonics for artificial intelligence and neuromorphic computing. *Nat. Photonics* **15**, 102–114 (2021).

10. Capmany, J. & Novak, D. Microwave photonics combines two worlds. *Nat. Photonics* **1**, 319–330 (2007).

11. Marpaung, D., Yao, J. & Capmany, J. Integrated microwave photonics. *Nat. Photonics* **13**, 80–90 (2019).

12. McMahon, P. L. The physics of optical computing. *Nat. Rev. Phys.* **5**, 717–734 (2023).

13. Marković, D., Mizrahi, A., Querlioz, D. & Grollier, J. Physics for neuromorphic computing. *Nat. Rev. Phys.* **2**, 499–510 (2020).

14. Rashidinejad, A., Li, Y. & Weiner, A. M. Recent Advances in Programmable Photonic-Assisted Ultrabroadband Radio-Frequency Arbitrary Waveform Generation. *IEEE J. Quantum Electron.* **52**, 1–17 (2016).

15. Bogaerts, W. *et al.* Programmable photonic circuits. *Nature* **586**, 207–216 (2020).

16. Shu, H. *et al.* Microcomb-driven silicon photonic systems. *Nature* **605**, 457–463 (2022).

17. Zhang, M., Wang, C., Kharel, P., Zhu, D. & Lončar, M. Integrated lithium niobate electro-optic modulators: when performance meets scalability. *Optica* **8**, 652 (2021).

18. Xu, M. *et al.* Dual-polarization thin-film lithium niobate in-phase quadrature modulators for terabit-per-second transmission. *Optica* **9**, 61 (2022).

19. Patel, D., Samani, A., Veerasubramanian, V., Ghosh, S. & Plant, D. V. Silicon Photonic Segmented Modulator-Based Electro-Optic DAC for 100 Gb/s PAM-4 Generation. *IEEE Photonics Technol. Lett.* **27**, 2433–2436 (2015).

20. Dubé-Demers, R., LaRochelle, S. & Shi, W. Low-power DAC-less PAM-4 transmitter using a cascaded microring modulator. *Opt. Lett.* **41**, 5369 (2016).

21. Simard, A. D., Filion, B., Patel, D., Plant, D. & LaRochelle, S. Segmented silicon MZM for PAM-8 transmissions at 114 Gb/s with binary signaling. *Opt. Express* **24**, 19467 (2016).

22. Huynh, T. N. *et al.* Flexible Transmitter Employing Silicon-Segmented Mach–Zehnder Modulator With 32-nm CMOS Distributed Driver. *J. Light. Technol.* **34**, 5129–5136 (2016).

23. Samani, A. *et al.* Experimental parametric study of 128 Gb/s PAM-4 transmission system using a multi-electrode silicon photonic Mach Zehnder modulator. *Opt. Express* **25**, 13252 (2017).

24. Sobu, Y. *et al.* High-Speed Optical Digital-to-Analog Converter Operation of Compact Two-Segment All-Silicon Mach–Zehnder Modulator. *J. Light. Technol.* **39**, 1148–1154 (2021).

25. Jafari, O., Zhalehpour, S., Shi, W. & LaRochelle, S. DAC-Less PAM-4 Slow-Light Silicon Photonic Modulator Providing High Efficiency and Stability. *J. Light. Technol.* **39**, 5074–5082 (2021).

26. Wang, C. *et al.* Integrated lithium niobate electro-optic modulators operating at CMOS-compatible voltages. *Nature* **562**, 101–104 (2018).

27. Hu, Y. *et al.* On-chip electro-optic frequency shifters and beam splitters. *Nature* **599**, 587–593 (2021).

28. Hu, Y. *et al.* Integrated electro-optics on thin-film lithium niobate. Preprint at https://doi.org/10.48550/arXiv.2404.06398 (2024).

29. Zhang, M., Wang, C., Cheng, R., Shams-Ansari, A. & Lončar, M. Monolithic ultra-high-Q lithium niobate microring resonator. *Optica* **4**, 1536 (2017).

30. Zhu, X. *et al.* Twenty-nine million intrinsic $Q$-factor monolithic microresonators on thin-film lithium niobate. *Photonics Res.* **12**, A63 (2024).

31. Zhou, H. *et al.* Photonic matrix multiplication lights up photonic accelerator and beyond. *Light Sci. Appl.* **11**, 30 (2022).

32. Cundiff, S. T. & Weiner, A. M. Optical arbitrary waveform generation. *Nat. Photonics* **4**, 760–766 (2010).

33. Wang, J. *et al.* Reconfigurable radio-frequency arbitrary waveforms synthesized in a silicon photonic chip. *Nat. Commun.* **6**, 5957 (2015).

34. Tan, M. *et al.* Photonic RF Arbitrary Waveform Generator Based on a Soliton Crystal Micro-Comb Source. *J. Light. Technol.* **38**, 6221–6226 (2020).

35. Fischer, B. *et al.* Autonomous on-chip interferometry for reconfigurable optical waveform generation. *Optica* **8**, 1268 (2021).

36. Cox, C. H., Ackerman, E. I., Betts, G. E. & Prince, J. L. Limits on the performance of RF-over-fiber links and their impact on device design. *IEEE Trans. Microw. Theory Tech.* **54**, 906–920 (2006).

37. Kharel, P., Reimer, C., Luke, K., He, L. & Zhang, M. Breaking voltage–bandwidth limits in integrated lithium niobate modulators using micro-structured electrodes. *Optica* **8**, 357 (2021).

38. Xue, S. *et al.* Full-spectrum visible electro-optic modulator. *Optica* **10**, 125 (2023).

39. Renaud, D. *et al.* Sub-1 Volt and high-bandwidth visible to near-infrared electro-optic modulators. *Nat. Commun.* **14**, 1496 (2023).

40. He, L. *et al.* Low-loss fiber-to-chip interface for lithium niobate photonic integrated circuits. *Opt. Lett.* **44**, 2314 (2019).

41. Botez, D., Garrod, T. & Mawst, L. J. High CW wallplug efficiency 1.5 micron-emitting diode lasers. in *2015 IEEE Photonics Conference (IPC)* 551–552 (IEEE, Reston, VA, 2015). doi:10.1109/IPCon.2015.7323726.

42. Mashanovitch, M. *et al.* High-Power, Efficient DFB Laser Technology for RF Photonics Links. in *2018 IEEE Avionics and Vehicle Fiber-Optics and Photonics Conference (AVFOP)* 1–2 (IEEE, Portland, OR, 2018). doi:10.1109/AVFOP.2018.8550469.

43. Jin, W. *et al.* Hertz-linewidth semiconductor lasers using CMOS-ready ultra-high-Q microresonators. *Nat. Photonics* **15**, 346–353 (2021).

44. Yu, M. *et al.* Integrated electro-optic isolator on thin-film lithium niobate. *Nat. Photonics* **17**, 666–671 (2023).

45. Xiang, C. *et al.* 3D integration enables ultralow-noise isolator-free lasers in silicon photonics. *Nature* **620**, 78–85 (2023).

46. Guo, X. *et al.* High-performance modified uni-traveling carrier photodiode integrated on a thin-film lithium niobate platform. *Photonics Res.* **10**, 1338 (2022).

47. Xu, X. *et al.* 11 TOPS photonic convolutional accelerator for optical neural networks. *Nature* **589**, 44–51 (2021).

48. Nahmias, M. A. *et al.* Photonic Multiply-Accumulate Operations for Neural Networks. *IEEE J. Sel. Top. Quantum Electron.* **26**, 1–18 (2020).

49. Shen, Y. *et al.* Deep learning with coherent nanophotonic circuits. *Nat. Photonics* **11**, 441–446 (2017).

50. Gupta, S., Agrawal, A., Gopalakrishnan, K. & Narayanan, P. Deep Learning with Limited Numerical Precision. *Proc. 32nd Int. Conf. Mach. Learn.* **37**, 1737–1746 (2015).

51. Hamerly, R., Bernstein, L., Sludds, A., Soljačić, M. & Englund, D. Large-Scale Optical Neural Networks Based on Photoelectric Multiplication. *Phys. Rev. X* **9**, 021032 (2019).

52. Chen, Z. *et al.* Deep learning with coherent VCSEL neural networks. *Nat. Photonics* **17**, 723–730 (2023).

53. Lin, Z. *et al.* 120 GOPS Photonic tensor core in thin-film lithium niobate for inference and in situ training. *Nat. Commun.* **15**, 9081 (2024).

54. Ou, S. *et al.* Hypermultiplexed Integrated Tensor Optical Processor. Preprint at https://doi.org/10.48550/arXiv.2401.18050 (2024).

55. Hu, Y. *et al.* Integrated lithium niobate photonic computing circuit based on efficient and high-speed electro-optic conversion. Preprint at https://doi.org/10.48550/arXiv.2411.02734 (2024).

56. Hu, Y. *et al.* High-efficiency and broadband on-chip electro-optic frequency comb generators. *Nat. Photonics* **16**, 679–685 (2022).

57. Song, Y., Hu, Y., Zhu, X., Yang, K. & Lončar, M. Octave-spanning Kerr soliton frequency combs in dispersion- and dissipation-engineered lithium niobate microresonators. *Light Sci. Appl.* **13**, 225 (2024).

58. Bai, B. *et al.* Microcomb-based integrated photonic processing unit. *Nat. Commun.* **14**, 66 (2023).

59. Chen, Z. *et al.* Efficient erbium-doped thin-film lithium niobate waveguide amplifiers. *Opt. Lett.* **46**, 1161 (2021).

60. Cardenas, J. *et al.* Wide-bandwidth continuously tunable optical delay line using silicon microring resonators. *Opt. Express* **18**, 26525 (2010).

61. Koenig, S. *et al.* Wireless sub-THz communication system with high data rate. *Nat. Photonics* **7**, 977–981 (2013).

62. Pfeifle, J. *et al.* Coherent terabit communications with microresonator Kerr frequency combs. *Nat. Photonics* **8**, 375–380 (2014).

63. Marin-Palomo, P. *et al.* Microresonator-based solitons for massively parallel coherent optical communications. *Nature* **546**, 274–279 (2017).

64. Rizzo, A. *et al.* Massively scalable Kerr comb-driven silicon photonic link. *Nat. Photonics* **17**, 781–790 (2023).

65. Ghelfi, P. *et al.* A fully photonics-based coherent radar system. *Nature* **507**, 341–345 (2014).

66. Zhu, S. *et al.* Integrated lithium niobate photonic millimeter-wave radar. Preprint at https://doi.org/10.48550/arXiv.2311.09857 (2023).

67. Liu, W. *et al.* A fully reconfigurable photonic integrated signal processor. *Nat. Photonics* **10**, 190–195 (2016).

68. Feng, H. *et al.* Integrated lithium niobate microwave photonic processing engine. *Nature* **627**, 80–87 (2024).

**Acknowledgements:** The authors thank Kyle Richard, Neil Hoffman, Mark Roberts, and Keysight Technologies, Inc., for technological support, Jeremiah Jacobson for assistance, Prof. David Plant and Prof. Gage Hills for discussions.

**Funding:** This work is supported by Korea Advanced Institute of Science and Technology (KAIST) NRF-2022M3K4A1094782, Defense Advanced Research Projects Agency (DARPA) HR001120C0137, National Science Foundation (NSF) OMA-2137723, NSF 2138068, NSF EEC-1941583, Department of the Navy N6833522C0413, Amazon Web Services (AWS) A50791, DRS Daylight Solutions, Inc. Award A56097, NASA 80NSSC22K0262, NIH P41EB015903, NIH R21EY031895, and Singapore National Research Foundation (NRF2022-QEP2-01-P07, NRF-NRFF15-2023-0005). Y.S. acknowledges support from the AWS Generation Q Fund at the Harvard Quantum Initiative. L.Ma. acknowledges support from the Capes-Fulbright and Behring foundation fellowships. H.W. acknowledges support from the National Science Foundation Graduate Research Fellowship Program.

**Competing interests:** F.Y., D.R., M.Z., and M.Lo. are involved in developing lithium niobate technologies at HyperLight Corporation. The authors declare no other competing interests.

**Data and materials availability:** All data needed to evaluate the conclusions in the paper are present the paper and/or the Supplementary Information

**Disclaimer:** The views, opinions and/or findings expressed are those of the authors and should not be interpreted as representing the official views or policies of the Department of Defense or the U.S. Government.

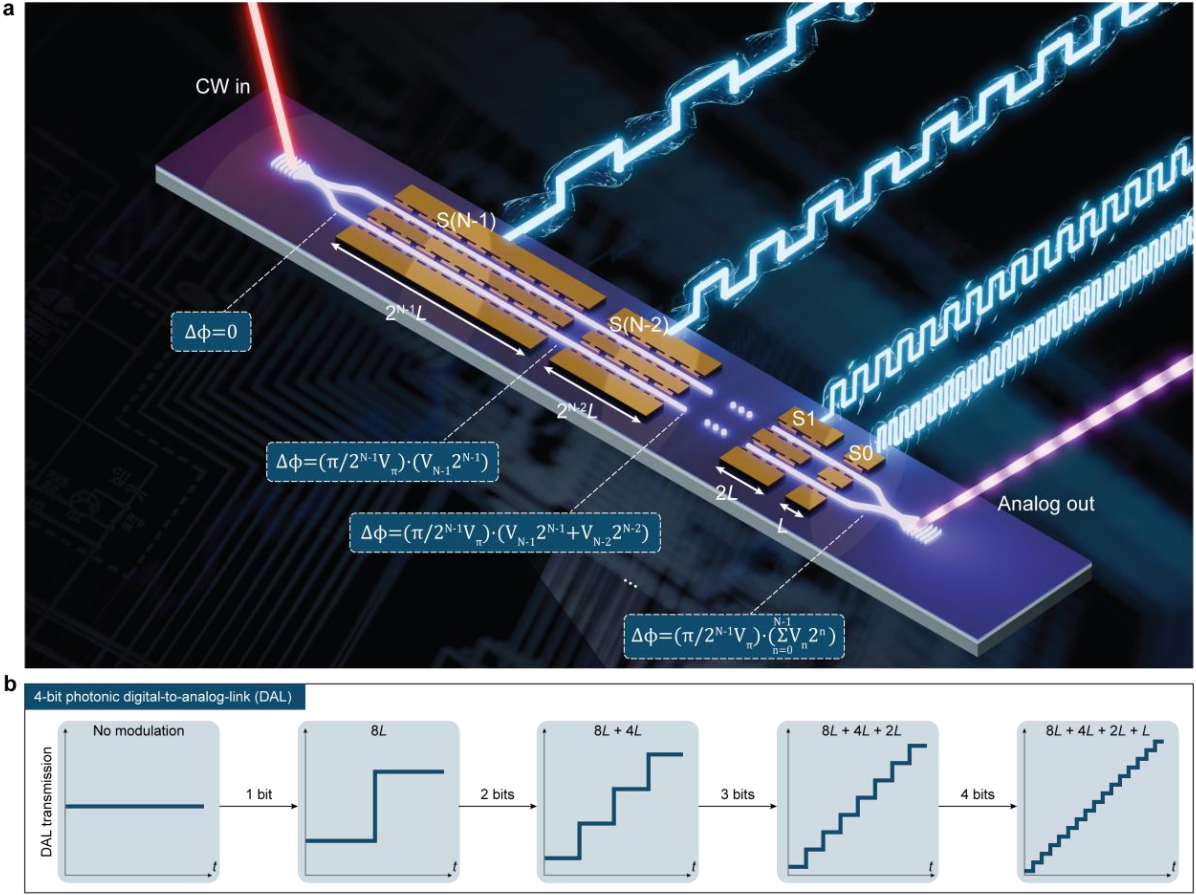

**Fig. 1 | Photonic-integrated EO-DiAL concept. a**, Illustration of an *N* bit resolution EO-DiAL architecture. Key components include an MZI featuring *N* modulation segments with lengths (hence $V_\pi$s) related by factors of two. The shortest (longest) segment controls the least (most) significant bit of the *N*-bit analog value (see Eq. 1 and derivation in the Methods section). The $n^{\text{th}}$ bit is set by a digital electronic input $V_n \in \{V_{high}, V_{low}\}$ applied to the $n^{\text{th}}$ segment with length $2^n L$. Segments are labeled by the bit they address: the least significant bit (bit 0) corresponds to the shortest segment and is labeled S0; the most significant bit (bit $N-1$) corresponds to the longest segment and is labeled S($N$-1). The insets mark the phase accumulation $\Delta\phi$ after each segment, from longest to shortest segment. Since $\Delta\phi$ is bit-wise addressable, the MZI transmission is as well, when operating near quadrature point. Thus, a fast switching sequence of $V_n(t)$ amounts to a continuously and rapidly updating *N*-bit-resolved analog data stream output by the DAL device. Notably, DAC occurs directly from the electronic to optical domain, without the overhead of electronic digital-to-analog converters. **b**, 4-bit resolved transmission of the EO-DiAL implemented on TFLN, in this work.

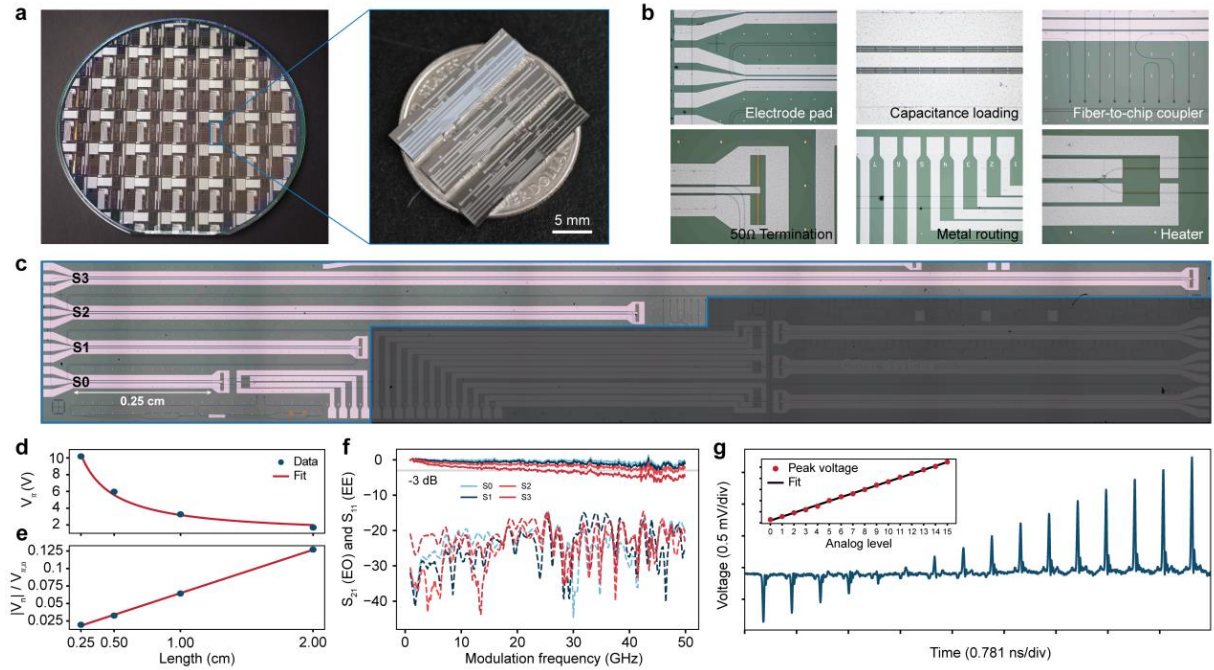

**Fig. 2 | EO-DiAL implementation and device characterization a,** Wafer-scale process (left) implemented for the EO-DiAL devices (right) used in this work. **b,** Individual components comprising the EO-DiAL device, including electrode pads, capacitance-loaded RF transmission lines, fiber-to-chip grating couplers, 50Ω terminators, metal routing, and heaters. For detailed information, see Methods section. **c,** Optical microscope image of EO-DiAL architecture implemented with TFLN nanophotonics, consisting of four successive modulation segments (S0-S3) with lengths 0.25, 0.50, 1.00, and 2.00 cm, respectively. **d,** Measured $V_\pi$ for each modulation segment. **e,** Ratio $|V_n|/V_{\pi,n}$, for $|V_n|$ used to calibrate deviations from factor-of-two $V_\pi$ relationship due to fabrication imperfection. **f,** Electro-optic forward transmission (EO $S_{21}$) and electric-electric input reflection (EE $S_{11}$) for each modulation segment. Segment S3 achieves an EO 3-dB bandwidth of 30 GHz, while nearly flat bandwidth responses up to 50 GHz are maintained for segments S0-S2. EE return loss is low for all segments. **g,** Linear optical intensity relation across all 16 levels encoded within the dynamic range of the 4-bit EO-DiAL (21.505 ps per level). Inset shows linear fit of optical intensity vs. analog level (0-15, or four bits of information).

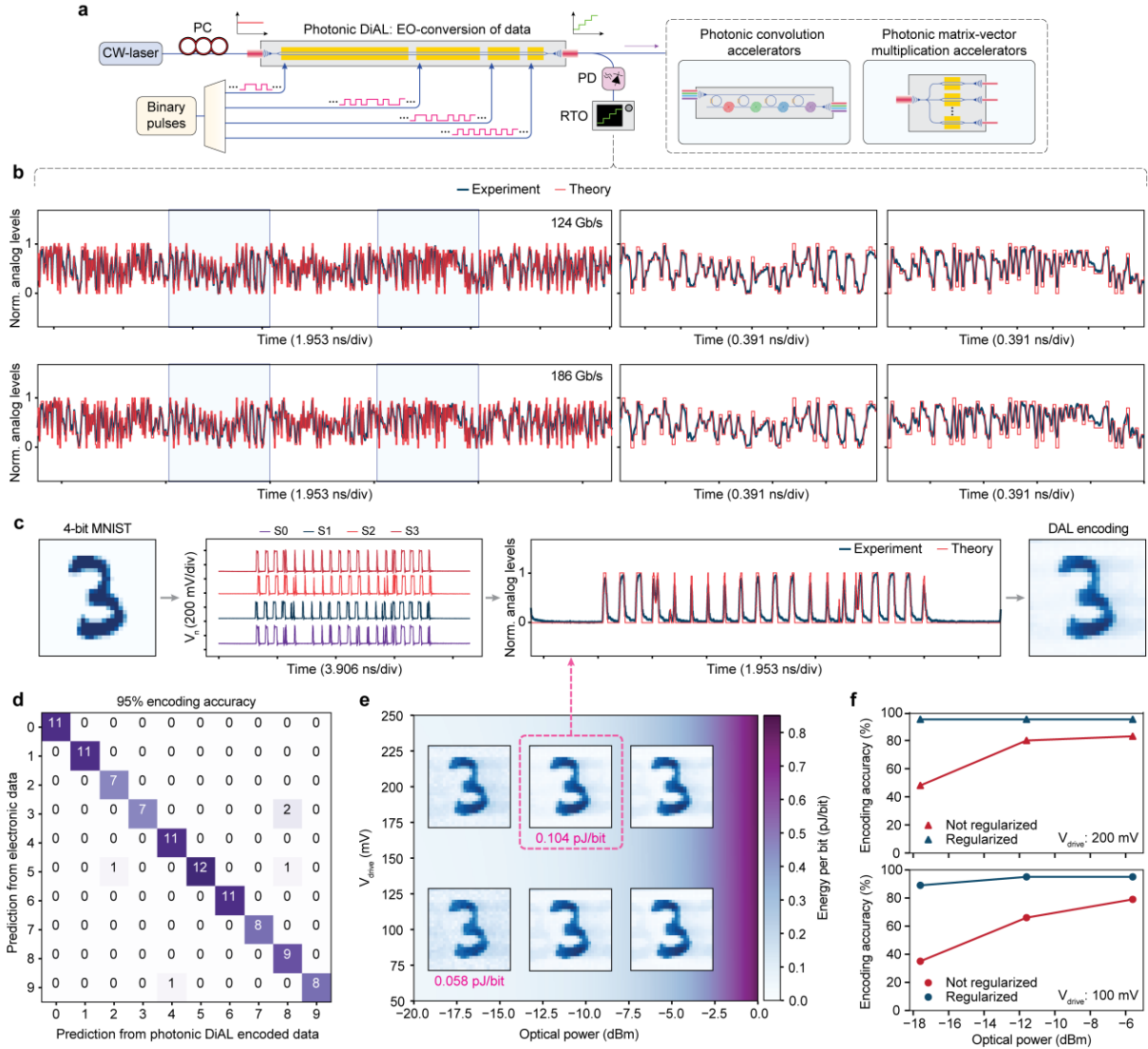

**Fig. 3 | High-speed and energy efficient EO conversion of data. a,** Schematic of the EO-DiAL operated as an EO digital-to-analog converter in photonic computing systems. **b,** A random sequence of 500 4-bit words encoded by the device, at effective data rates of 124 Gb/s (top) and 186 Gb/s (bottom), corresponding to independent modulation rates of 31 Gb/s and 46.5 Gb/s per electrode segment. Right panels show zoom-ins of the start and end of the traces. **c,** Left to right: down-sampled MNIST image (28 by 28 pixels) flattened into 4-bit pixel arrays of size 784 each; binary modulation patterns applied to segments S0-S3; EO conversion and DAC by the device at 186 Gb/s and 0.104 pJ/bit; and reconstructed MNIST image from photodetected optical intensities. A weak shadowing effect is due to the bandwidth limits in the system, corresponding to finite rise and drop times. **d,** Confusion matrix comparing classification outcomes between applying a machine learning model on test sets consisting of (i) original MNIST (defined as ground truth) and (ii) EO-DiAL-encoded MNIST, using experimental parameters identical to **c**. 95% of EO-DiAL-encoded images faithfully reproduce the fully electronic computations and thus are considered accurately encoded. **e,** Encoding quality and **f,** encoding accuracy as a function of electronic and optical energy consumption. The background color gradient includes all optical and electrical energy consumptions, as well as microheater dissipation when thermally biasing at quadrature. With 100 mV for each $|V_n|$ and -17.5 dBm optical power incident on the detector, amounting to 0.058 pJ/bit, 89% of images are accurately encoded.

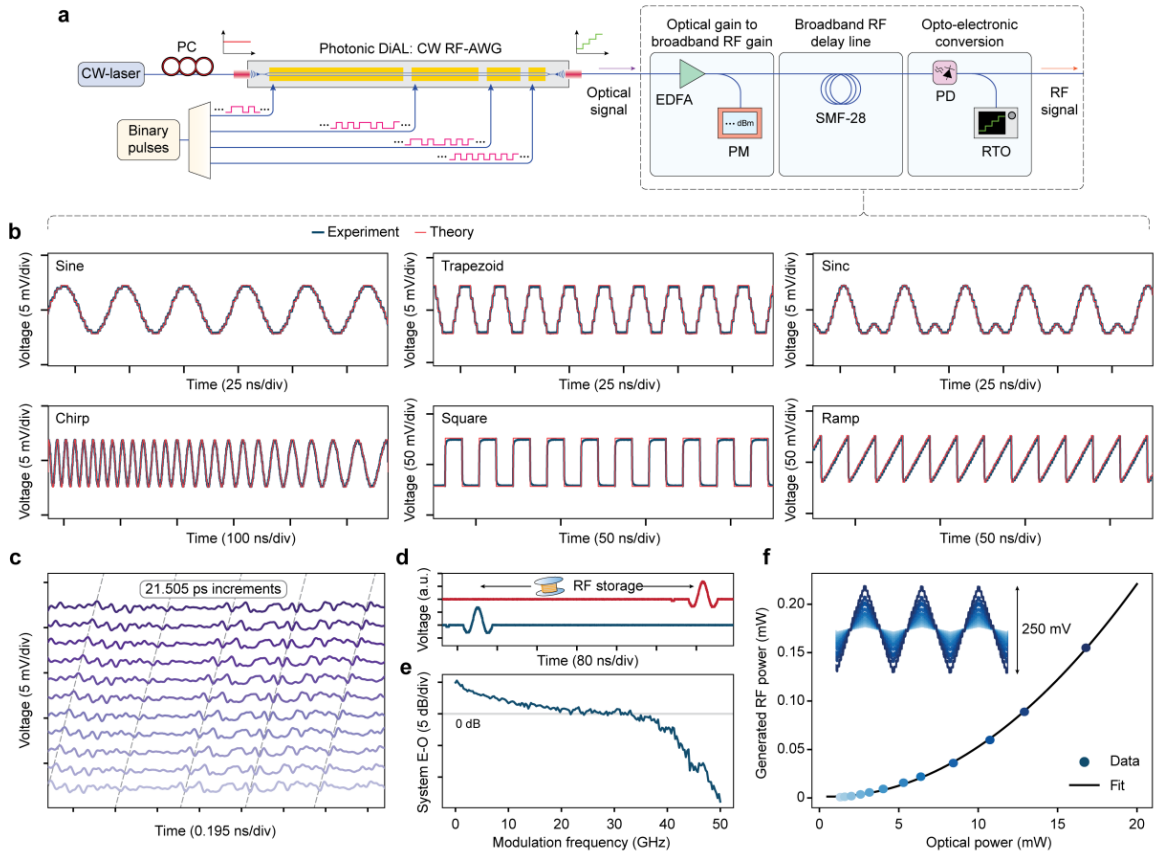

**Fig. 4 | Microwave-photonic RF-AWG. a,** Schematic of the EO-DiAL operated for microwave-photonic RF-AWG based on CW optical input. Optical gain and optical delays are subsequently applied for the broadband optical-microwave gain transfer, delay, storage and release of the RF arbitrary waveform. **b,** Standard waveforms encoded by the EO-DiAL onto the optical carrier and then converted into the electronic domain using a fast photodetector. **c,** Tuning of optical path length (in this case, incremental delays of about 21.505 ps) prior to demodulation enables broadband, word-by-word delay of a random RF waveform, as well as **d,** broadband storage (prolonged delay of about 480 ns) and release of an RF pulse. **e,** Broadband digital-to-analog gain on the RF arbitrary waveform via amplification of both the optical carrier and modulation sidebands followed by OE conversion. The full system response (EO $S_{21}$) surpasses the total digital electronic input power (-6 dBm, indicated by 0 dB level), up to 35 GHz modulation frequency, owing to 11.6 dB of optical gain and photodetector transimpedance gain (750 V/W) applied. **f,** Estimated generated RF power of triangular waves as a function of average optical power incident on the photodiode. This RF power is purely based on the voltage induced by photocurrent across an $R$=50 Ω-load. The black line corresponds to a fitted conversion photodiode gain of about 19.4 V/W.